%% file: ICRC2023_EHE.tex
\documentclass[a4paper,11pt]{article}

\usepackage{pos}
\usepackage{lipsum} 
\usepackage[separate-uncertainty]{siunitx}
\DeclareSIUnit\years{yrs}
\DeclareSIUnit\photoelectron{PE}
\usepackage[section]{placeins}

\title{Search for Extremely High Energy Neutrinos \\ with IceCube}

\ShortTitle{Search for EHE $\nu$ with IceCube}

\author{The IceCube Collaboration \\{\normalsize \normalfont(a complete list of authors can be found at the end of the proceedings)}\\}

\emailAdd{mmeier@icecube.wisc.edu}
\emailAdd{brianclark@icecube.wisc.edu}

\abstract{
Extremely high energy (EHE) neutrinos (with energies above $10^7$ GeV) are produced in interactions of the highest energy cosmic rays. A primary contribution to the EHE neutrino flux is expected from so-called cosmogenic neutrinos produced when ultra high energy cosmic rays interact with ambient photon backgrounds. Observations of these EHE neutrinos with IceCube can probe the nature of cosmic rays beyond the energies for resonant photo-pion production (GZK cutoff). We present a new event selection of extremely high energy neutrinos by more effectively identifying and rejecting high multiplicity muon bundles with respect to previous analyses. Furthermore, we show the expected improvements of the quasi-differential upper limits on the EHE neutrino flux using 12 years of IceCube data.

\vspace{4mm}
{\bfseries Corresponding authors:}
Maximilian Meier$^{1*}$, Brian A. Clark$^{2}$\\
{$^{1}$ \itshape Chiba University}\\
{$^{2}$ \itshape University of Maryland, College Park}\\[4mm]
$^*$ Presenter

\ConferenceLogo{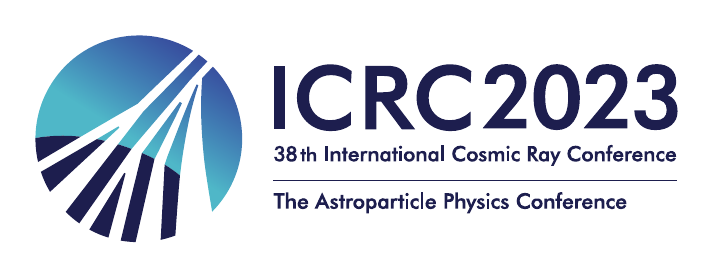}

\FullConference{The 38th International Cosmic Ray Conference (ICRC2023)\\ 26 July -- 3 August, 2023\\ Nagoya, Japan}
}

\begin{document}

\maketitle

\section{Introduction}\label{sec1}

Extremely-high-energy neutrinos (EHE, $\gtrsim 10^7$\,GeV) are unique messengers to the violent, high-redshift universe. At these high energies, other Standard Model messengers do not arrive from great distances ($\sim 50$\,Mpc). Cosmic rays are deflected by magnetic fields, and above $\sim 10^{19.5}$\,eV they are expected to interact with the Cosmic Microwave Background via the famous Greisen-Zatsepin-Kuz'min (GZK) effect~\cite{Greisen1966, Zatsepin:1966jv}. Gamma-rays are also expected to be attenuated via interactions with the CMB and Extragalactic Background Light; they can additionally be absorbed by dust in sources, in so-called ``Compton-thick" environments. 
Unlike cosmic-rays and gamma-rays, neutrinos are uncharged and interact only through the Weak force. This allows them to travel through space undeflected and unattenuated. A UHE flux of neutrinos is expected to arise ``in-flight" from the aforementioned GZK interactions (``cosmogenic" neutrinos), but also from the environments immediately surrounding the astrophysical accelerators themselves (``astrophysical" neutrinos).
The cosmogenic flux is expected to encode unique and complementary information about the ultra-high-energy cosmic-ray (UHECR) flux. In particular, the shape and normalization of the neutrino flux is expected to encode information about the chemical composition, redshift evolution, and maximum accelerating energy of cosmic ray accelerators.

The IceCube Neutrino Observatory has previously searched for this flux of EHE neutrinos. In this proceeding, we report on efforts to expand this experimental search. This revision contains an additional 5.5 years of detector livetime as well as updates to the event selection which improve the efficiency of the analysis, especially in the Southern Sky.

\section{Detector, Data, and Simulation}
The IceCube Neutrino Observatory is a cubic-kilometer neutrino detector built at the South Pole~\cite{IceCube:2016zyt}. IceCube is composed of 5,160 Digital Optical Modules (DOMs) buried in the glacier at South Pole, Antarctica. The DOMs consist of a downward-facing photomultiplier tube with digitization and readout electronics. They are distributed along 86 ``strings" between depths of 1500 and 2500m, with 60 DOMs to a string spaced 15m apart vertically. Each string is $\sim125$\,m apart laterally, resulting in a lattice of detectors instrumenting almost a gigaton of ice. IceCube underwent phased construction, reaching  22 strings (``IC22") by June 2007, and then 40, 59, 79, and finally 86 strings (``IC86") in 2008-2011 respectively.
In addition to this ``in-ice" component of IceCube, the detector also contains two pairs of tanks (each containing two DOMs) sitting above each string, composing the ``IceTop" instrument. IceTop enables measurements of the extensive air showers arising from cosmic ray interactions in the atmosphere.

IceCube observes neutrino interactions by looking for the Cherenkov light produced by neutrino-nucleon interactions in the ice. These Cherenkov photons are converted via the photoelectric effect in the PMT, and are observed in IceCube as charge ($Q$). IceCube can observe neutrinos of all flavors ($\nu_e$, $\nu_{\mu}$, $\nu_{\tau}$) and types ($\nu$, $\bar{\nu}$) primarily through two detection channels. At these very high energies, neutral-current (NC) interactions of all neutrino flavors, and charged-current interactions of $\nu_e$ produce spatially compact electromagnetic and hadronic showers, which are observed as roughly spherical depositions of light in the detector, or ``cascades". Charged-current interactions of $\nu_{\mu}$ and  $\nu_{\tau}$ produce long-lived daughter muon and tau leptons which can leave a long series of stochastic energy depositions in the detector, and therefore appear as ``tracks."
This second category of events is particularly important to the EHE search. It means that neutrinos can interact far outside of the IceCube detector volume, travel tens of kilometers and still arrive at the detector. This enlarges the detector's \textit{effective volume} by orders of magnitude relative to the cubic kilometer of \textit{instrumented volume}. In addition to astrophysical and cosmogenic neutrinos, IceCube also observes muons and neutrinos generated by cosmic ray interactions in the atmosphere. These ``atmospheric muons" and ``atmospheric neutrinos" are observed at a rate of $\sim3$\,kHz and a few mHz respectively, which both far exceed the \si{\micro\hertz} rate expected for astrophysical neutrinos, and \si{\nano\hertz} rate anticipated for cosmogenic neutrinos.

In this revision to the search for extremely high energy neutrinos, we analyze data from May 2010 to January 2023, covering data collected by IC79 and IC86 for a collective 12.36 years of livetime. The data from the detector has been completely recalibrated according to the ``Pass2" calibration campaign, which specifically leverages an updated and more accurate modeling of the DOM single-photoelectron charge response~\cite{IceCube:2020nwx}.
This is an addition of almost 5.5  years of new data relative to the previous search~\cite{EHE9yr}.
In contrast to this previous search, this iteration does not utilize data recorded before May 2010, corresponding to detector configurations predating IC79. This is because it is challenging to treat these earlier years consistently with the updated calibration techniques. The impact of removing these prior years is reduced, however, by their comparatively smaller effective areas and livetimes. We estimate the removal of IC22, IC40, and IC59 reduces the total integrated exposure of the detector to EHE neutrinos by no more than 10\%.

In order to design a search, we must simulate both signals (cosmogenic neutrinos) and backgrounds (astrophysical neutrinos, atmospheric neutrinos and muons). The atmospheric muon background is simulated using the \texttt{CORSIKA} air shower simulation framework~\cite{corsika} with the \texttt{SIBYLL2.3c} hadronic interaction model~\cite{sibyll2.3c} up to primary energies of \SI{e11}{\giga\electronvolt} assuming a primary cosmic ray flux from~\cite{gaisserh4a}. Neutrinos -- cosmogenic signal, as well as astrophysical and atmospheric neutrinos -- are simulated with \texttt{JULIET}~\cite{juliet} up to energies of \SI{e11}{\giga\electronvolt}. The baseline model for cosmogenic signal used here is~\cite{Ahlers_2010}. Astrophysical neutrinos are modeled based on previous IceCube measurements~\cite{northern_tracks_2022, cascades, hese_7}. For conventional and prompt atmospheric neutrinos models from~\cite{hkkms} and~\cite{berss} are used respectively.

\section{Event Selection}\label{sec2}

The main background for EHE neutrinos in IceCube are down-going high energy, high multiplicity muon bundles produced by cosmic ray air showers in the Earth's atmosphere. Additional backgrounds are astrophysical neutrinos as well as atmospheric neutrinos. 
The astrophysical neutrinos produce an isotropic flux of neutrinos reaching Earth with an almost equal predicted flavor ratio of $\nu_e:\nu_{\mu}:\nu_{\tau} = 1:1:1$, and a spectral index of about $\gamma \sim -2.5$, dominating the flux of atmospheric neutrinos at energies above \SI{100}{\tera\electronvolt}. Atmospheric neutrinos are produced in the same cosmic ray air showers as atmospheric muon bundles.
Their spectral index is with $\gamma \sim -3.7$ much softer than the spectral index of astrophysical neutrinos making them dominant at low energies.

The event selection approach is based on~\cite{EHE7yr}, where signal candidates are found by applying four consecutive steps aimed at reducing these atmospheric backgrounds.

\subsubsection*{Level 2}
In the fist step of the event selection only events with a total recorded charge of $Q_{\mathrm{tot}} \geq \SI{27500}{\photoelectron}$ and a number of hit DOMs of $n_{\mathrm{DOMs}} \geq 100$ are kept. This cut already rejects a large majority of atmospheric neutrinos, that are the dominant neutrino component at this stage.

\subsubsection*{Level 3: Track quality \label{subsec:l3}}
\begin{figure}
    \centering
    \includegraphics[width=1.\textwidth]{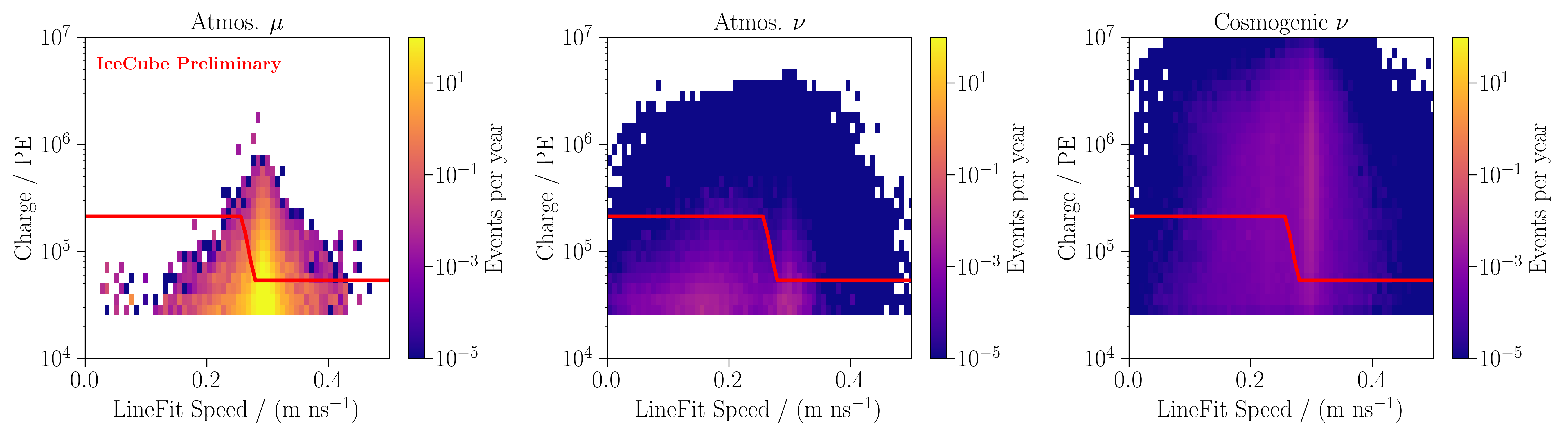}
    \caption{Distribution of LineFit Speed and charge $Q$ for atmospheric muons (left), atmospheric neutrinos (center), and cosmogenic neutrinos (right). The Level 3 track quality cut applied is shown as a red line.}
    \label{fig:L3_cut}
\end{figure}

The Level 3 cut is shown as the red line in Fig.~\ref{fig:L3_cut} for atmospheric muons, atmospheric neutrinos and cosmogenic neutrinos respectively. The cut is a two-dimensional cut in the plane of reconstructed event velocity (described later) and the total recorded charge $Q_{\mathrm{tot}}$. The Level 3 criterion has multiple purposes. It rejects atmospheric neutrinos, especially prompt neutrinos (see the middle panel), and also rejects mis-reconstructed atmospheric muon events and neutrino events. The velocity calculation relies on the ``LineFit" reconstruction algorithm~\cite{linefit}. The reconstruction assumes light traveling with a speed $\vec{v}$ along an infinite track. For a well reconstructed track the speed will be distributed closely around the speed of light. Cascades or mis-reconstructed tracks will have smaller reconstructed speeds. As a consequence, the LineFit speed can also be used to separate the final event sample into subsets of cascades and tracks, which is done at the pivot point of the Level 3 cut ($|| \vec{v} || = \SI{0.27}{\meter\per\nano\second}$).

\subsubsection*{Level 4: Muon bundle}
\begin{figure}
    \centering
    \includegraphics[width=.67\textwidth]{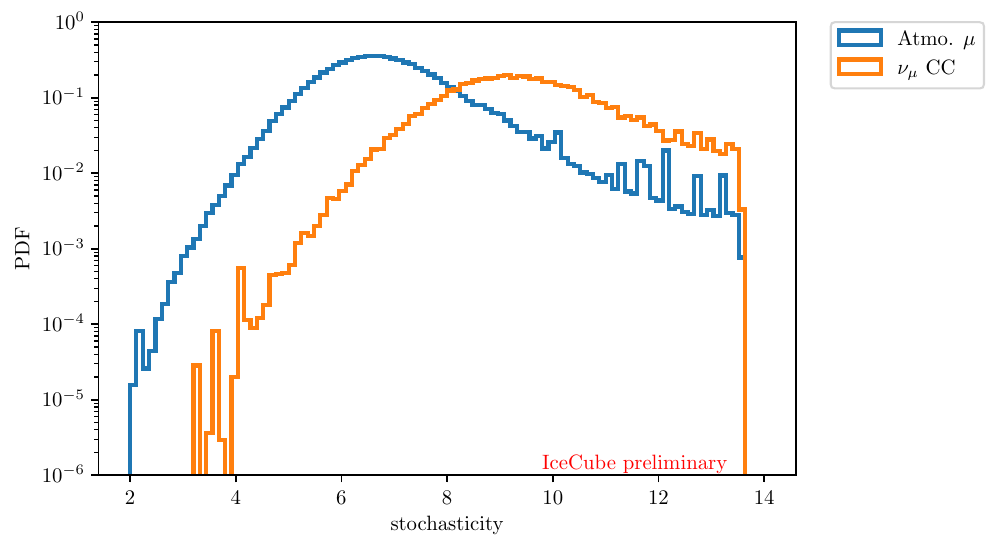}
    \caption{
        Distribution of stochasticity for atmospheric muons (muon bundles) and $\nu_{\mu} \, \mathrm{CC}$ (single muons) events. Smaller stochasticity values are more compatible with the muon bundle hypothesis.
    }
    \label{fig:stoch_classifier}
\end{figure}

\begin{figure}
    \centering
    \begin{minipage}{0.45\textwidth}
        \includegraphics[width=\textwidth]{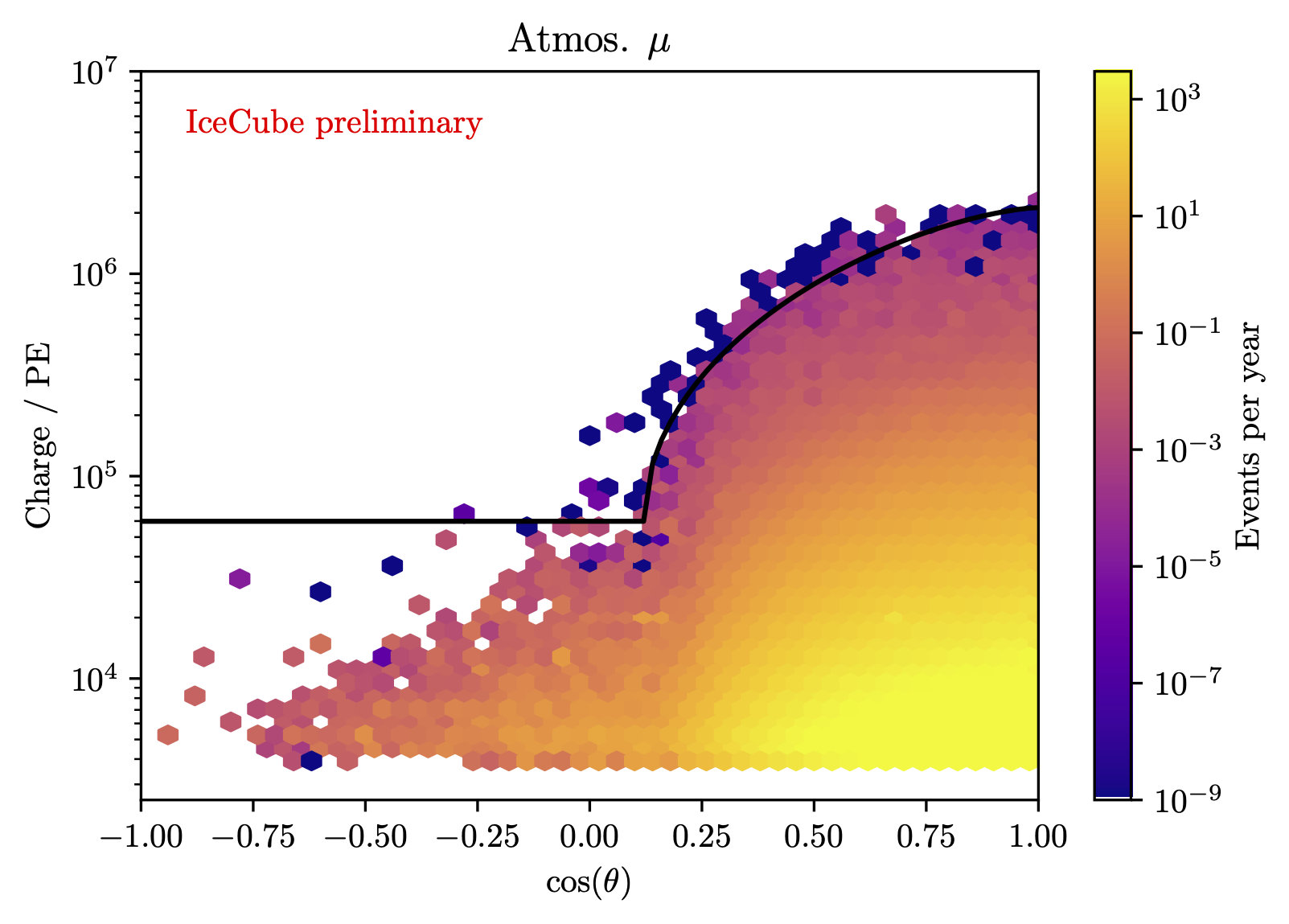}
    \end{minipage}
    \begin{minipage}{0.45\textwidth}
        \includegraphics[width=\textwidth]{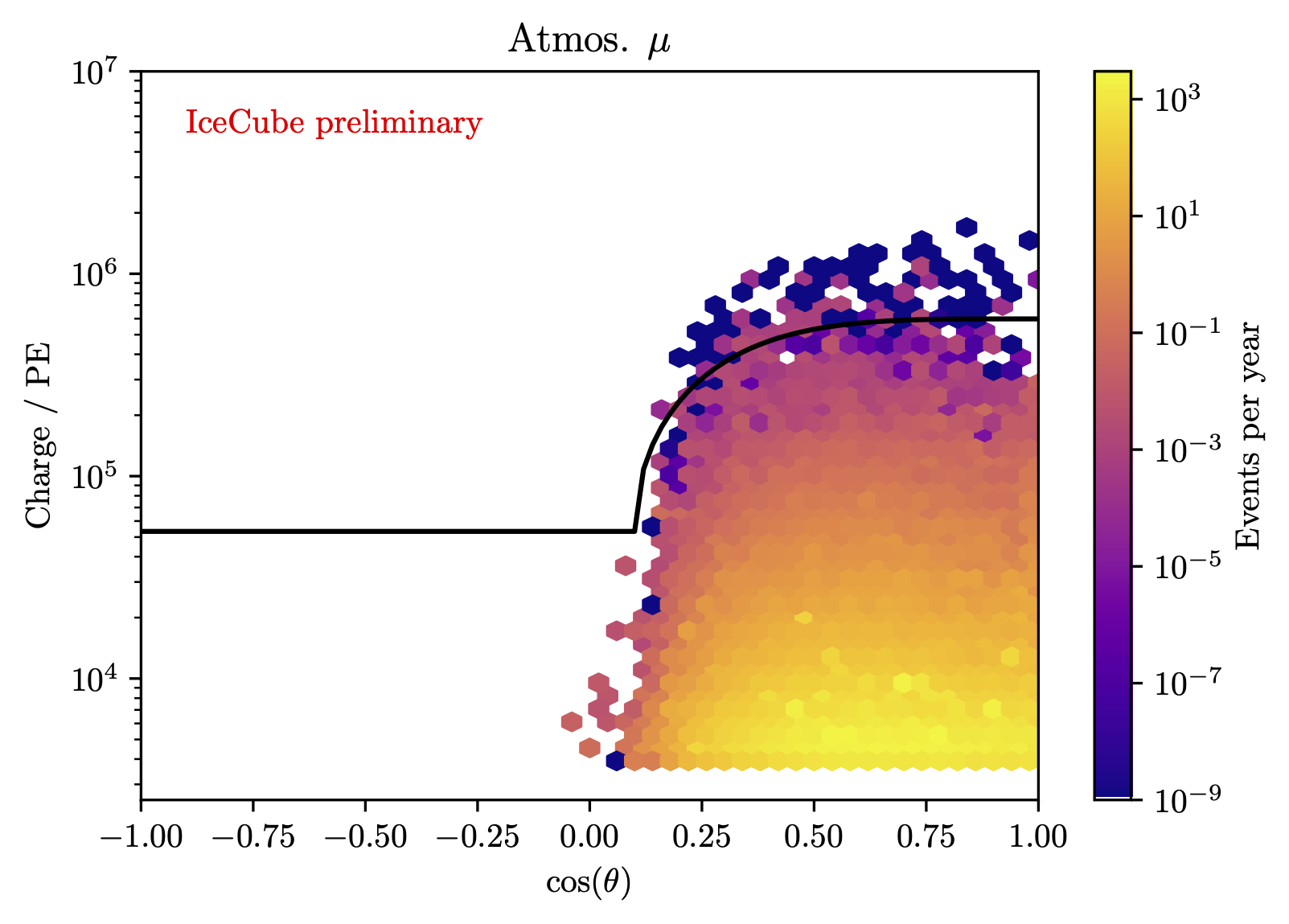}
    \end{minipage}
    \caption{Distribution of atmospheric muon background as a function of charge and $\cos(\theta)$ for small stochasticities (left) and large stochasticities (right). The events below the black lines are removed by the Level 4 criterion.}
    \label{fig:l4_cut}
\end{figure}

The goal of the Level 4 cut criterion is to remove the main background of down-going muon bundles. The cut is made in the 2D plane of reconstructed particle zenith $\cos(\theta)$ and total recorded charge $Q_{\mathrm{tot}}$ and is visible in Fig.~\ref{fig:l4_cut}. 
In this cut plane, the differences between signal (cosmogenic neutrinos) and dominant background (atmospheric muons) appears in both the zenith distribution and the energy loss profile of single muons/taus compared to muon bundles with large multiplicities. As the energy of a muon increases, its energy losses become more stochastic. In a muon bundle with the same total energy, the energy is distributed among many muons resulting in a superposition of lower energy muons losing their energy more continuously.
To obtain a measure of the "stochasticity" of an event, the energy loss profile is reconstructed using a segmented energy loss reconstruction~\cite{energy_reco_paper}. The reconstructed profile is then compared to a muon bundle PDF obtained with \texttt{PROPOSAL}~\cite{proposal} to get a stochasticity reconstruction: $\mathrm{stochasticity} = \sum_i \log (P(\Delta E_i/E)) / \mathrm{ndf}$, similar to a reduced log-likelihood. The distribution of the stochasticity is shown in Fig.~\ref{fig:stoch_classifier} for atmospheric muon background simulation and $\nu_{\mu} \, \mathrm{CC}$ events to represent single muons.

Fig.~\ref{fig:l4_cut} shows the 2D-distributions of events in charge and the cosine of the reconstructed zenith angle split into two stochasticity bins, with the boundary set at \num{8.37}. The solid black lines show the cuts applied to remove the majority of the atmospheric muon background. The split in stochasticity allows for a looser cut in the downgoing region $\cos(\theta) > 0$, increasing the signal efficiency in this step relative a version of the analysis without stocasticity binning.

\begin{figure}
    \centering
    \begin{minipage}{0.45\textwidth}
        \includegraphics[width=\textwidth]{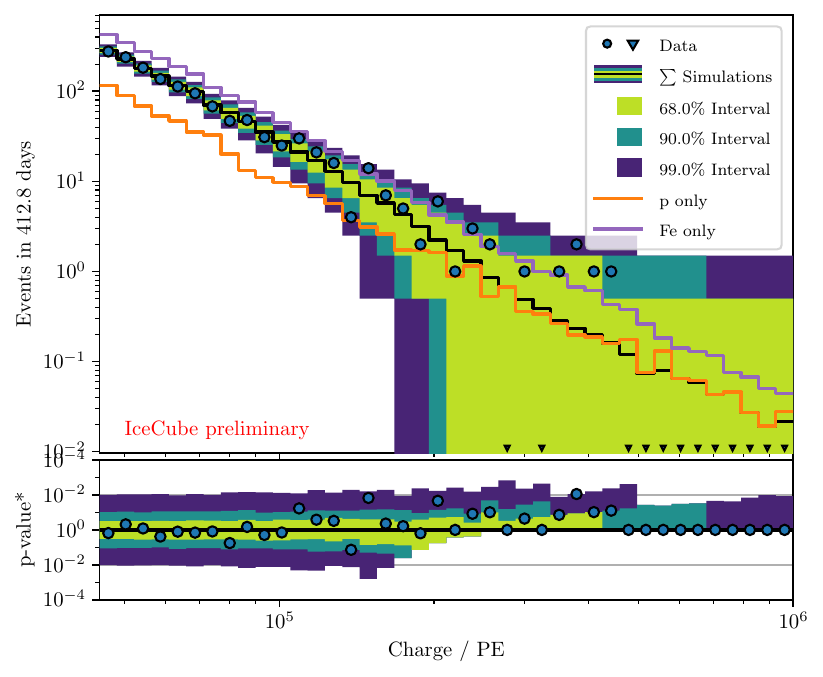}
    \end{minipage}
    \begin{minipage}{0.45\textwidth}
        \includegraphics[width=\textwidth]{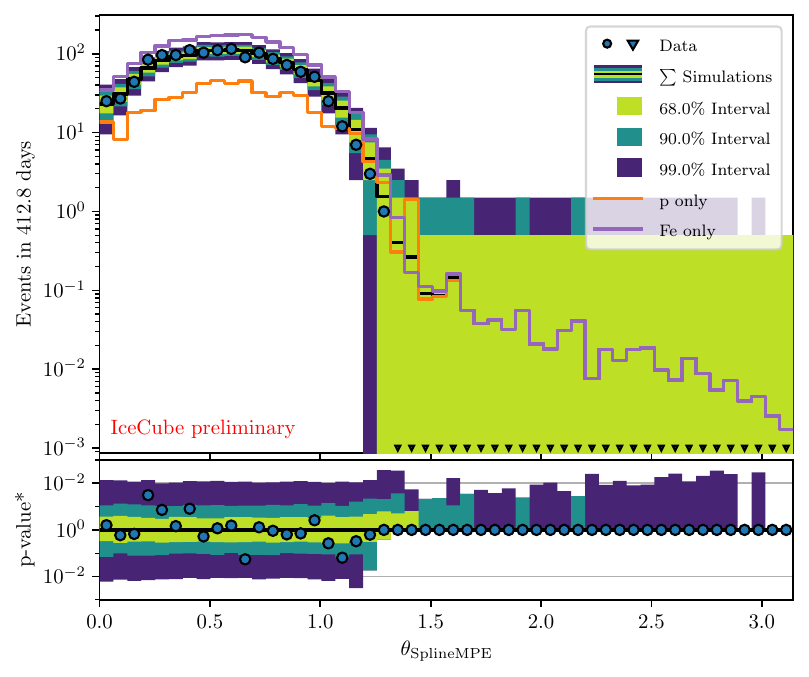}
    \end{minipage}
    \caption{Comparison between simulations and burnsample data before applying the Level 4 criterion. The simulations are weighted to a five component cosmic ray primary flux. Since the cosmic ray composition is not well known at the highest energies, a prediction assuming only protons/iron as cosmic ray primaries are also shown.}
    \label{fig:datamc}
\end{figure}

A comparison between simulations and burnsample data before applying the Level 4 criterion for the two main observables, charge and reconstructed zenith, is shown in Fig.~\ref{fig:datamc}.
We show these plots for a 10\% subsample (``burn sample") of the data, where we see excellent agreement between data and simulations in these two central cut variables.

\subsubsection*{Level 5: IceTop veto}

IceTop can be used to further reduce the background rate of atmospheric muons. IceTop hits correlated with an event in the in-ice detector can be found by extrapolating the reconstructed track to the surface and find the time $t_{\mathrm{CA}}$, where the track is at its closest approach to IceTop. Correlated IceTop hits are defined by the collections of hits that satisfy $\SI{-1}{\micro\second} \leq t_{\mathrm{CA}} \leq \SI{1.5}{\micro\second}$. Events are vetoed if they have two or more correlated hits in IceTop, reducing the remaining atmospheric muon background by about \SI{60}{\percent} but only reducing the all-sky neutrino rate by less than 5\%.

\begin{figure}
    \centering
    \begin{minipage}{0.49\textwidth}
        \includegraphics[width=1.\textwidth]{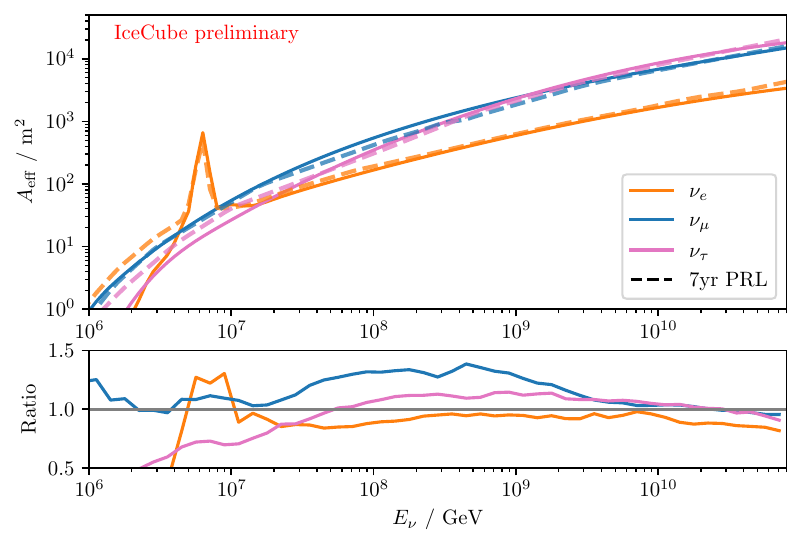}
        \caption{Zenith-averaged effective area for each neutrino flavor compared to the effective area for the event selection presented in~\cite{EHE7yr}.}
        \label{fig:effective_area}
    \end{minipage}
    \begin{minipage}{0.49\textwidth}
        \captionof{table}{Event expectations for different signal and background components after applying the IceTop veto. The expected astrophysical background is shown covering a range of spectral indices observed in other IceCube analyses. \label{tab:rates}}
        \centering
        \begin{tabular}{l||c|c}
            Flux component & Events in \SI{12}{\years} & $\gamma_{\mathrm{astro}}$ \\
            \hline
            $\mu_{\mathrm{atmo}}$ & 0.08 & \\
            $\nu_{\mathrm{conv}}$ & 0.17 & \\
            $\nu_{\mathrm{prompt}}$ & 0.03 & \\
            $\nu_{\mathrm{astro}, \mathrm{tracks}}$ \hfill \cite{northern_tracks_2022} & 8.62 & \num{-2.37} \\
            $\nu_{\mathrm{astro}, \mathrm{cascades}}$ \hfill \cite{cascades} & 4.78 & \num{-2.53}\\
            $\nu_{\mathrm{astro}, \mathrm{HESE}}$ \hfill \cite{hese_7} & 1.20 & \num{-2.89} \\
            \hline
            $\nu_{\mathrm{GZK}}$ \hfill \cite{Ahlers_2010} & 5.35 & \\
        \end{tabular}
    \end{minipage}
\end{figure}

The zenith-averaged neutrino effective area for the event selection (before applying the IceTop veto) is shown in Fig.~\ref{fig:effective_area} compared to the event selection presented in~\cite{EHE7yr}. The new event selection mostly improved the $\nu_{\mu}$ effective area between \SI{10}{\peta\electronvolt} and \SI{1}{\exa\electronvolt} by about 30\%, while reducing the $\nu_e$ and $\nu_{\tau}$ effective area between \SI{1}{\peta\electronvolt} and \SI{10}{\peta\electronvolt} to reduce the background of astrophysical neutrinos.
Expected event rates for different components and flux assumptions for \num{12}~years of IceCube data are listed in Tab.~\ref{tab:rates}.

\section{GZK Model Tests and Differential Limit}
After applying the event selection the sample is split into a subset of tracks and cascades as described in Sec.~\ref{subsec:l3}. The energy and arrival direction is reconstructed for both sub-samples using a likelihood-based reconstruction with a track and a cascade hypothesis respectively. A binned Poisson likelihood approach is used to fit to the data (following~\cite{EHE9yr}):
\begin{equation}
    \mathcal{L} (\lambda_{\mathrm{GZK}}, \lambda_{\mathrm{astro}}) = \prod_i P(n_i | \lambda_{\mathrm{GZK}} \mu_{i, \mathrm{GZK}} + \lambda_{\mathrm{astro}} \mu_{i, \mathrm{astro}} + \mu_{i, \mathrm{bkg}}),
    \label{eq:likelihood}
\end{equation}
where the two free parameters are the relative normalization to the signal GZK model $\lambda_{\mathrm{GZK}}$, and the relative normalization of the astrophysical nuisance flux $\lambda_{\mathrm{astro}}$.

The compatibility of the data with different GZK models will be tested using a likelihood-ratio test
\begin{equation}
    \Lambda = \log \left( \frac{\mathcal{L}(\hat{\lambda}_{\mathrm{GZK}}, \hat{\lambda}_{\mathrm{astro}})}{\mathcal{L}(\lambda_{\mathrm{GZK}} = 1, \hat{\hat{\lambda}}_{\mathrm{astro}})} \right).
\end{equation}

\begin{figure}
    \centering
    \includegraphics[width=1.\textwidth]{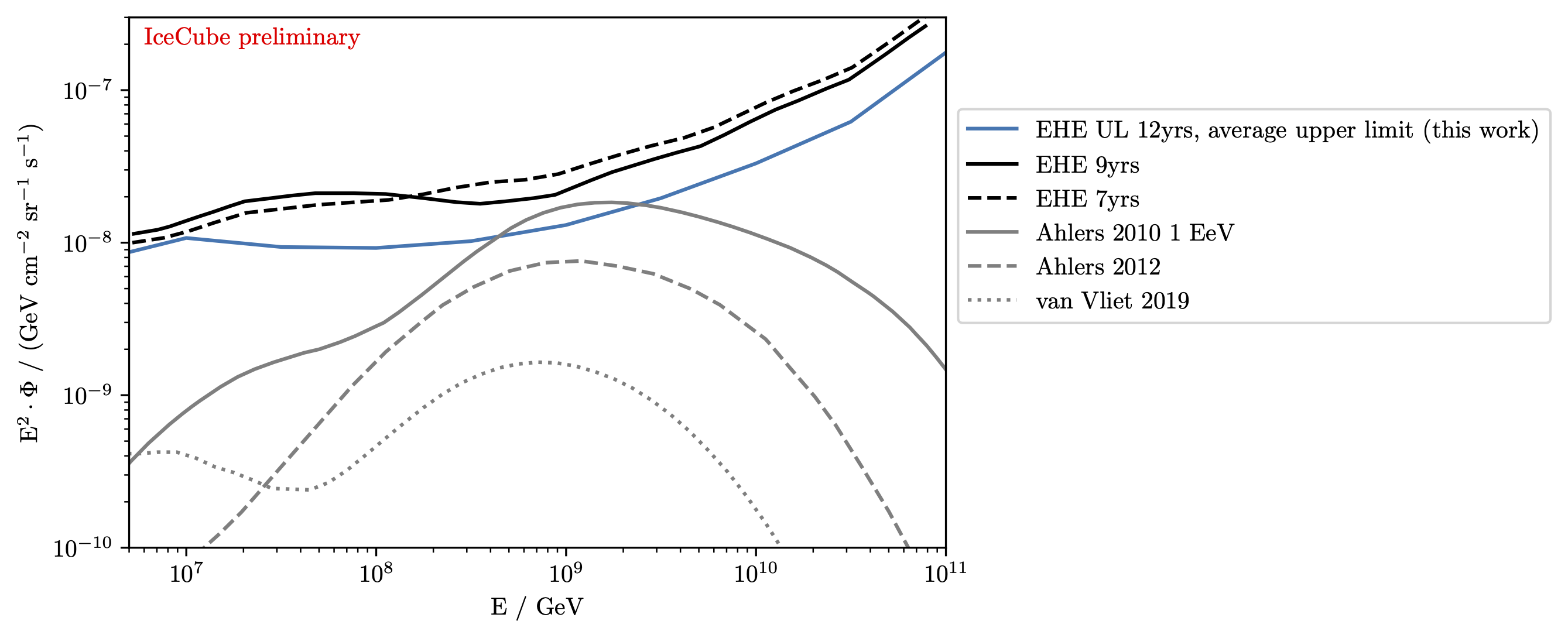}
    \caption{Sensitivity to the \SI{90}{\percent} CL differential upper limit on the ultra high energy neutrino flux for neutrinos energies between \SI{3}{\peta\electronvolt} and \SI{100}{\exa\electronvolt}. The expected average upper limit is compared to previous upper limits set by IceCube~\cite{EHE7yr, EHE9yr} and to cosmogenic neutrino flux models~\cite{Ahlers_2010, Ahlers_2012, van_Vliet_2019}. The model shown from~\cite{van_Vliet_2019} assumes $\gamma = 2.5$, $E_{\mathrm{max}} = \SI{e20}{\electronvolt}$, $m = 3.4$ and a \SI{10}{\percent} proton fraction.}
    \label{fig:differential_limit}
\end{figure}

A differential limit can also be constructed to obtain a more model independent constraint on the UHE neutrino flux. The differential limit is constructed in the same way as described in~\cite{EHE9yr}. The same likelihood formalism (Eq.~\ref{eq:likelihood}) is used, but for each tested energy $E_c$ an $E^{-1}$ signal flux with a width of one energy decade centered around $E_c$ is injected. Then, for each $E_c$ a Feldman-Cousins \SI{90}{\percent} confidence interval is constructed~\cite{Feldman_1998}.
The expected average upper limit to the differential UHE neutrino flux (using the most conservative background expectation from~\cite{northern_tracks_2022} for the astrophysical neutrino flux) is compared to previous IceCube results in Fig.~\ref{fig:differential_limit}.

\section{Conclusion}\label{sec3}

In this contribution, we have described work towards an updated search for Extremely High Energy (EHE) neutrinos with the IceCube detector. The search leverages new selection variables which improve the overall efficiency of the analysis and adds almost 5.5 years of additional detector livetime. The new search will have world-leading sensitivity to the flux of neutrinos at PeV energies and above.

\bibliographystyle{ICRC}
\bibliography{references}

\clearpage

\input{authorlist_IceCube.tex}

\end{document}

%% file: authorlist_IceCube.tex
\section*{Full Author List: IceCube Collaboration}

\scriptsize
\noindent
R. Abbasi$^{17}$,
M. Ackermann$^{63}$,
J. Adams$^{18}$,
S. K. Agarwalla$^{40,\: 64}$,
J. A. Aguilar$^{12}$,
M. Ahlers$^{22}$,
J.M. Alameddine$^{23}$,
N. M. Amin$^{44}$,
K. Andeen$^{42}$,
G. Anton$^{26}$
% ...
\\
\\
% ...
$^{12}$ Universit{\'e} Libre de Bruxelles, Science Faculty CP230, B-1050 Brussels, Belgium \\
% ...
$^{17}$ Department of Physics, Loyola University Chicago, Chicago, IL 60660, USA \\
$^{18}$ Dept. of Physics and Astronomy, University of Canterbury, Private Bag 4800, Christchurch, New Zealand \\
% ...
$^{22}$ Niels Bohr Institute, University of Copenhagen, DK-2100 Copenhagen, Denmark \\
$^{23}$ Dept. of Physics, TU Dortmund University, D-44221 Dortmund, Germany \\
% ...
$^{26}$ Erlangen Centre for Astroparticle Physics, Friedrich-Alexander-Universit{\"a}t Erlangen-N{\"u}rnberg, D-91058 Erlangen, Germany \\
% ...
$^{40}$ Dept. of Physics and Wisconsin IceCube Particle Astrophysics Center, University of Wisconsin{\textendash}Madison, Madison, WI 53706, USA \\
% ...
$^{42}$ Department of Physics, Marquette University, Milwaukee, WI, 53201, USA \\
% ...
$^{44}$ Bartol Research Institute and Dept. of Physics and Astronomy, University of Delaware, Newark, DE 19716, USA \\
% ...
$^{63}$ Deutsches Elektronen-Synchrotron DESY, Platanenallee 6, 15738 Zeuthen, Germany  \\
$^{64}$ Institute of Physics, Sachivalaya Marg, Sainik School Post, Bhubaneswar 751005, India \\
% ...

\subsection*{Acknowledgements}

\noindent
The authors gratefully acknowledge the support from the following agencies and institutions:
USA {\textendash} U.S. National Science Foundation-Office of Polar Programs,
U.S. National Science Foundation-Physics Division,
U.S. National Science Foundation-EPSCoR,
Wisconsin Alumni Research Foundation,
Center for High Throughput Computing (CHTC) at the University of Wisconsin{\textendash}Madison,
Open Science Grid (OSG),
Advanced Cyberinfrastructure Coordination Ecosystem: Services {\&} Support (ACCESS),
Frontera computing project at the Texas Advanced Computing Center,
U.S. Department of Energy-National Energy Research Scientific Computing Center,
Particle astrophysics research computing center at the University of Maryland,
Institute for Cyber-Enabled Research at Michigan State University,
and Astroparticle physics computational facility at Marquette University;
Belgium {\textendash} Funds for Scientific Research (FRS-FNRS and FWO),
FWO Odysseus and Big Science programmes,
and Belgian Federal Science Policy Office (Belspo);
Germany {\textendash} Bundesministerium f{\"u}r Bildung und Forschung (BMBF),
Deutsche Forschungsgemeinschaft (DFG),
Helmholtz Alliance for Astroparticle Physics (HAP),
Initiative and Networking Fund of the Helmholtz Association,
Deutsches Elektronen Synchrotron (DESY),
and High Performance Computing cluster of the RWTH Aachen;
Sweden {\textendash} Swedish Research Council,
Swedish Polar Research Secretariat,
Swedish National Infrastructure for Computing (SNIC),
and Knut and Alice Wallenberg Foundation;
European Union {\textendash} EGI Advanced Computing for research;
Australia {\textendash} Australian Research Council;
Canada {\textendash} Natural Sciences and Engineering Research Council of Canada,
Calcul Qu{\'e}bec, Compute Ontario, Canada Foundation for Innovation, WestGrid, and Compute Canada;
Denmark {\textendash} Villum Fonden, Carlsberg Foundation, and European Commission;
New Zealand {\textendash} Marsden Fund;
Japan {\textendash} Japan Society for Promotion of Science (JSPS)
and Institute for Global Prominent Research (IGPR) of Chiba University;
Korea {\textendash} National Research Foundation of Korea (NRF);
Switzerland {\textendash} Swiss National Science Foundation (SNSF);
United Kingdom {\textendash} Department of Physics, University of Oxford.